\documentclass[aps,twocolumn,showpacs,superscriptaddress,floatfix,nofootinbib]{revtex4-2}

\pdfoutput=1
\usepackage[utf8]{inputenc}
\usepackage[english]{babel}
\usepackage[T1]{fontenc}
\usepackage{csquotes}
\usepackage{physics}

\usepackage{amsmath,amsfonts}

\newcommand{\hilb}[0]{\mathcal{H}}

\usepackage[dvipsnames]{xcolor}

\usepackage[colorlinks=true]{hyperref}
\usepackage[capitalise]{cleveref}

\usepackage{listings}
\usepackage{parskip}

\usepackage{graphicx}

\usepackage{qcircuit}

\begin{document}
\title{Variational dynamics as a ground-state problem on a quantum computer}

\author{Stefano Barison}
\email{stefano.barison@epfl.ch}
\affiliation{Institute of Physics, \'{E}cole Polytechnique F\'{e}d\'{e}rale de Lausanne (EPFL), CH-1015 Lausanne, Switzerland}
\affiliation{National Centre for Computational Design and Discovery of Novel Materials MARVEL, EPFL, Lausanne, Switzerland}

\author{Filippo Vicentini}
\affiliation{Institute of Physics, \'{E}cole Polytechnique F\'{e}d\'{e}rale de Lausanne (EPFL), CH-1015 Lausanne, Switzerland}

\author{Ignacio Cirac}
\affiliation{Max-Planck-Institut f\"ur Quantenoptik, Hans-Kopfermann-Str. 1, 85748 Garching, Germany}

\author{Giuseppe Carleo}
\affiliation{Institute of Physics, \'{E}cole Polytechnique F\'{e}d\'{e}rale de Lausanne (EPFL), CH-1015 Lausanne, Switzerland}
\affiliation{National Centre for Computational Design and Discovery of Novel Materials MARVEL, EPFL, Lausanne, Switzerland}

\begin{abstract}
We propose a variational quantum algorithm to study the real time dynamics of quantum systems as a ground-state problem. 
The method is based on the original proposal of Feynman and Kitaev to encode time into a register of auxiliary qubits.
We prepare the Feynman-Kitaev Hamiltonian acting on the composed system as a qubit operator and find an approximate ground state using the Variational Quantum Eigensolver.
We apply the algorithm to the study of the dynamics of a transverse field Ising chain with an increasing number of spins and time steps, proving a favorable scaling in terms of the number of two qubit gates.
Through numerical experiments, we investigate its robustness against noise, showing that the method can be use to evaluate dynamical properties of quantum systems and detect the presence of dynamical quantum phase transitions by measuring Loschmidt echoes. 
\end{abstract}

\maketitle

\section{Introduction}
\label{sec:intro}

Recent years have seen tremendous developments in the fabrication, control, and deployment of quantum computing systems, and it is now possible to access quantum computing platforms with up to a hundred of qubits on the cloud \cite{ibm_services,google_services,wright2019_natc,ionq_services}.
These devices are expected to surpass the capabilities of classical computers in specific tasks such as factorization \cite{Shor94Book}, database search \cite{grover96book} and quantum simulation \cite{Santoro2006,Kassal2008,georgescu14qs}.

Among these tasks, quantum simulation was the first envisioned application of quantum computers \cite{Feynman1982} and has been proved to be of polynomial complexity (BQP) on devices making use of quantum resources \cite{lloyd1996,abrams97qs}.
Due to its potential impact on many different areas of physics \cite{cade2020_prb,arute2020_arxiv}, chemistry \cite{Cao2019_cr,bauer2020_cr} and materials science \cite{sheng2021quantum,bassman2021simulating}, significant amounts of research have been devoted to such application.

However, the realisation of an universal quantum simulator remains far in the future due to combined effect of limited connectivity and noisy gates: the noise sets a maximum number of operations that can be performed without affecting the fidelity, and the sparse connection graph among qubits increases the total number of physical operations that must be performed to implement an algorithm.
As an example, several recent works describing optimized algorithms for quantum dynamics require a number of quantum operations (gates) well beyond the possibility of current hardware to surpass classical capabilities \cite{childs2018_pnas,babbush2018_prx,yunseong2019_npjqi,motta2021_npjqi}.
A computational strategy that works around the constraints of the hardware is the use of hybrid quantum-classical algorithms \cite{Peruzzo_2014,Ying2017rte,Motta_2019,ollitrault2019quantum,Yuan2019tv,cerezo2020variational,barison2021efficient,Hsuan_2021,Barratt_2021}.
In this approach, the quantum computer is used to perform a subroutine of limited depth, while the whole algorithm is governed by a classical computer.
In the context of quantum dynamics, several hybrid algorithms approximating the real-time evolution of a physical quantum system have been proposed \cite{Ying2017rte,Yuan2019tv,otten2019noiseresilient,C_rstoiu_2020,commeau2020variational,bharti2020quantum,benedetti2020hardwareefficient,barison2021efficient,Hsuan_2021,Barratt_2021}. 
Those algorithms rely on a variational trial state (ansatz) whose parameters $\theta $ are optimised classically in order to approximate the desired state. 

A remarkably different approach, that has not been throughly investigated on quantum computers yet, treats time as a quantum degree of freedom by encoding it into an auxiliary system called the \textit{clock}.
In that picture, the state $\ket{\psi(t)}$ at a certain time $t$ is represented as the composite state $\ket{\psi(t)}\otimes\ket{t}$. 
This approach is of interest because it allows to encode the dynamics of a system into a static superposition of such states, $\sum_t \ket{\psi(t)}\otimes\ket{t}$, and each state $\ket{\psi(t)}$ can be obtained by projecting the \textit{clock} into the corresponding state $\ket{t}$.
Given a time-dependent system, it is possible to construct an Hamiltonian for the joint physical-clock system whose ground-state encodes the whole evolution of the original system. 
Such construction, originally proposed by Feynman and Kitaev \cite{Feynman85,kitaev02_qc,Caha_2018,Bausch_2018} in the case of a discrete clock and later extended to continuous time \cite{Giovannetti2015PRDQuantumTime}, is effectively recasting the problem of quantum dynamics into a ground-state problem.
This perspective leads to a different classical variational principle \cite{McClean_2013,Tempel_2014,McClean_2015}, that allows the application of ground-state techniques from quantum chemistry and condensed-matter physics to quantum dynamics.
A similar construction was also recently proposed to address the simulation of time-dependent Hamiltonians \cite{Watkins_2022}.

In this work we extend the original idea to perform the variational optimisation of the ground state with an hybrid quantum algorithm.
First, we show how to prepare the Feynman-Kitaev Hamiltonian efficiently as a qubit operator.
Then, we consider a variational ansatz for the ground-state and optimise the circuit parameters using the Variational Quantum Eigensolver (VQE) \cite{Peruzzo_2014,McClean_2016,Kandala2017}.
We reiterate that this approach allows us to study quantum dynamics using a wide number of techniques originally developed for ground-state problems.
Moreover, it is related to the McLachlan's variational principle \cite{McLachlan1964,McClean_2013}, which is suggested to be most consistent variational principle for quantum simulation \cite{Yuan2019tv,barison2021efficient}.
Finally, once a variational approximation of the ground-state is obtained, we propose to use it to evaluate dynamical properties of the physical system and detect dynamical quantum phase transitions \cite{heyl2018_dpt_review}. 

The structure of this paper is as follows: in Section \ref{sec:methods} we present the Feynman-Kitaev Hamiltonian and its efficient mapping onto a series of qubit operators, while in Section \ref{sec:results} we apply the method to the study of a Transverse Field Ising chain, comparing the results to both exact and noisy Trotter simulations.
\Cref{sec:discussion} concludes the paper with some considerations and outlooks on the proposed algorithm.

\section{Methods}
\label{sec:methods}

In this section we will describe how to construct the Feynman-Kitaev Hamiltonian given a time-evolving physical system and how to apply it in a  hybrid quantum algorithm.

\subsection{Feynman-Kitaev Hamiltonian}
\label{sec:feynman-kitaev-hamiltonian}

We consider a quantum system governed by an Hamiltonian $\hat{H}$ acting on the physical Hilbert space $\hilb_p$.
To simplify the discussion, we assume that $\hat{H}$ is time-independent, but that is not a requirement.
Given the initial state of the system $|\psi(t=0)\rangle$, the state at every time $t$ is $ e^{-i \hat{H} t}\ket{\psi(0)} = \hat{U}(t,0)\ket{\psi(0)}$, formal solution of the time-dependent Schr\"{o}dinger equation.

The unitary transformations $\hat{U}(t,t') = \hat{U}(t-t')$ form a one-parameter group, therefore we can use the group composition property to write

\begin{equation}
    \label{eq:unitary_dec}
     \hat{U}(t,0)\ket{\psi(0)} = \hat{U}(t,t_{n-1})\dots \hat{U}(t_2,t_1)\hat{U}(t_1,0) \ket{\psi(0)} \, ,
\end{equation}

where we supposed $t= t_{n}>t_{n-1}>\dots>t_1>t_0=0$. 
By taking $t_{i}-t_{i-1} = dt  \quad \forall i \in \{1,\dots, n\} $, the expression in \cref{eq:unitary_dec} becomes $\hat{U}^{n}(dt)\ket{\psi(0)}$, namely we have discretised the time evolution of the system in $n$ equal sub-steps of length $dt$.

We then consider an auxiliary \textit{clock} system $\hilb_c = \mathrm{span}\{\ket{0},\ket{1}, \dots \ket{n}\}$ with $\braket{i}{j} = \delta_{ij} \quad \forall i,j \in \{0,\dots, n\}$.
They correspond to the different times $\{t_0, t_1, \dots t_n\}$. 
We then define the \textit{history state} to be the coherent superposition of the physical states $\ket{\psi(t)}$ at different times in the joint physical-clock system $\hilb = \hilb_p \otimes \hilb_c$,
\begin{equation}
\label{eq:hist_state}
    \ket{\Psi} = \frac{1}{\sqrt{n+1}}\sum_{i=0}^{n} \hat{U}^{i}(dt)\ket{\psi(0)} \ket{i} = \frac{1}{\sqrt{n+1}}\sum_{i=0}^{n}\ket{\psi_i} \ket{i} \, ,
\end{equation}
where we defined for simplicity $\hat{U}^{i}(dt)\ket{\psi(0)} =\ket{\psi_i}$.
From the history state it is possible to extract the physical state at any time $t$ by projecting on the auxiliary qubits

\begin{equation}
    \ket{\psi_i} = \sqrt{n+1} \braket{i}{\Psi} \, .
\end{equation}

We now define the Feynman-Kitaev Hamiltonian $\hat{C}$ as the operator whose ground state corresponds to the history state with energy 0 \cite{Feynman1982,kitaev02_qc}. Similarly to \cite{McClean_2013}, we present it in a form amenable to implementation on quantum computers. 
The Hamiltonian can be split into three terms:

\begin{equation}
\label{eq:clock_hami}
    \hat{C}   = \hat{C}_0 + \frac{1}{2}( \hat{C}_1 - \hat{C}_2)  \, , \\
\end{equation}

where

\begin{equation}
\begin{aligned}
  \hat{C}_0 &= \big[ \hat{\mathbf{I}} - \dyad{\psi(0)}{\psi(0)} \big] \otimes \dyad{0}{0} \, ,\\
  \hat{C}_1 &= \sum_{i=0}^{n-1} \hat{\mathbf{I}} \otimes \dyad{i}{i} + \hat{\mathbf{I}} \otimes \dyad{i+1}{i+1}  \, ,\\
  \hat{C}_2 &= \sum_{i=0}^{n-1} \hat{U}(dt)\otimes \dyad{i+1}{i} + \hat{U}^{\dagger}(dt) \otimes \dyad{i}{i+1}  \, .\\
\end{aligned}
\end{equation}

$\hat{C}_0$ favours the initial state $\ket{\psi(0)}$ by giving a positive contribution to the energy for any other state.
$\hat{C}_1$ and $\hat{C}_2$ encode for the evolution of the physical and auxiliary quantum system through time, by giving a positive contribution to the energy when the state at different times does not evolve according to the unitary operator $\hat{U}(dt)$.
We will call $ \hat{U}(dt)\otimes \dyad{i+1}{i}$ the forward unitary and its adjoint the backward unitary.
By construction, we have that the energy $E(\ket{\psi}) = \bra{\psi} \hat{C} \ket{\psi}\geq 0 $ for every state $\ket{\psi}$ and the equality holds only for $\ket{\psi} = \ket{\Psi}$.

Considering a state $\ket{\phi} = \sqrt{1-\epsilon^2}\ket{\Psi} + \epsilon\ket{\delta}$ with $\epsilon\approx 0,\, \, \braket{\Psi}{\delta} = 0 $ so that it's close to the target state, it is possible to show that the infidelity  between $\ket{\phi}$ and the target $\ket{\Psi}$ is $1-F(\phi, \Psi) = \epsilon^2$, where we defined $F(\phi, \Psi) =  |\langle \phi | \Psi \rangle |^2$, given  $\ket{\phi}$ and $\ket{\Psi}$ pure states. 
Meanwhile, the energy is given by $E(\ket{\phi}) \approx k\epsilon^2$ for some constant $k \geq E_1$, where $E_1$ is the energy of the first excited state of the Feynman-Kitaev Hamiltonian.
The computationally-costly infidelity is therefore upper-bounded by the normalized energy $1-F(\phi, \Psi) \leq E(\ket{\phi})/E_1$, that can therefore be used as a convergence metric in a variational optimisation.

\begin{figure*}[htb]
       \includegraphics[width=1.0\textwidth]{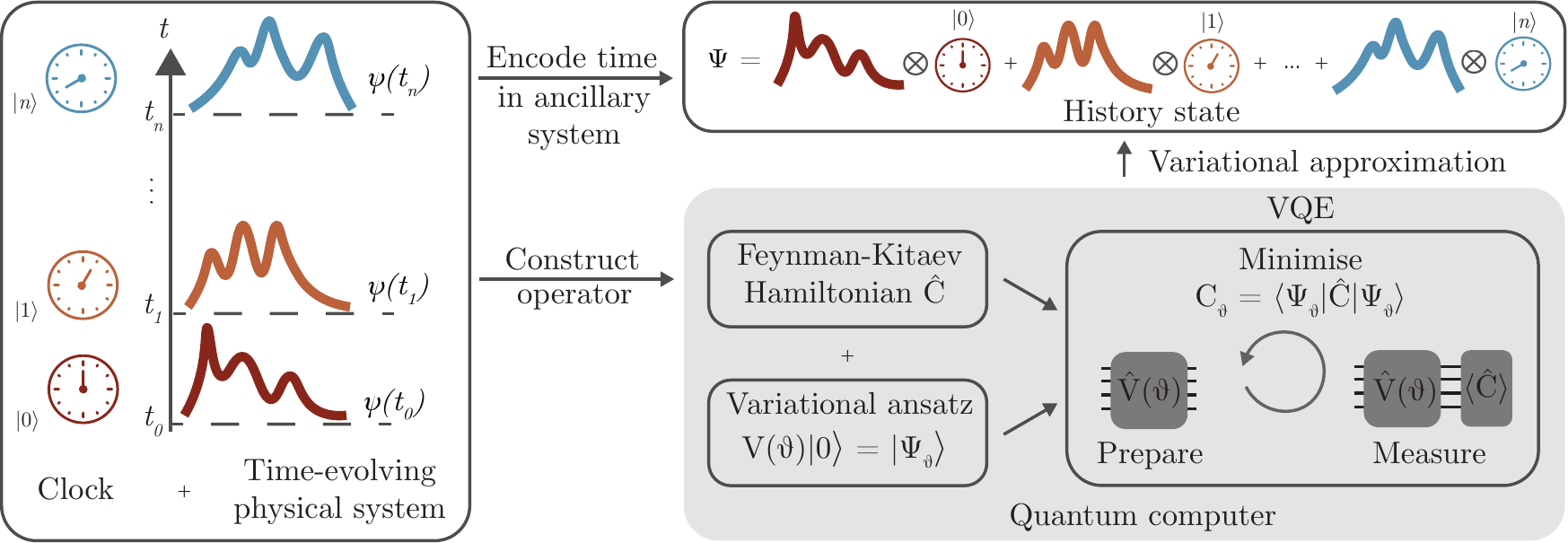} 
        \caption{Sketch of the VFK method. Coupling the physical system to an auxiliary clock system we can encode an entire time evolution into a time-independent state. We prepare a variational approximation of the history state by measuring the expectation value of Feynman-Kitaev Hamiltonian and its derivatives on quantum computer.}
        \label{fig:vfk_sketch}
\end{figure*}

\subsection{Efficient mapping on Quantum hardware}
\label{sec:mapping-hamiltonian-on-hardware}

We want to encode the history state on the qubits of a quantum computer, therefore we need to find an efficient mapping for the Feynman-Kitaev Hamiltonian onto qubit operators.
The physical and auxiliary subsystems are encoded into a different partition of qubits.
For the physical system, we will resort to one of the many established mappings \cite{Jordan93,Bravyi02_bk}.
The auxiliary register, instead, is a $n+1$-level system and there are in principle combinatorially many ways to map it to qubits \cite{Sawaya_20}.
In this work we considered both the standard binary and the Gray mapping \cite{gray53} to encode every state $\ket{i}$ into a bit string state.
Since $n_a$ binary digits have $2^{n_a}$ different values, encoding $n$ different time steps requires $n_{a} = \log_{2}(n)$ auxiliary qubits using those mappings.
The main difference among binary and Gray encoding is  how many (qu-)bits must be flipped when going from one time-step $\ket{i}$ to the next.
Under binary encoding one will eventually need to flip all qubits, such as when going from $\ket{i=3}\leftrightarrow \ket{011}$ to $\ket{i=4} \leftrightarrow \ket{100}$ in the case of $n_a = 3$.
Gray code, instead, is built in order to flip only one bit  at every step, and therefore the time-step operator $\dyad{i}{i+1}$ can be implemented with a single qubit rotation on 1 qubit.

Consider the time step $\ket{i}$ and its Gray binary mapping $\ket{x_{n_a}\dots x_2x_1}$, $x_i \in \{0,1\} \, \, \forall i$.
Let $x_j$ be the difference bit between $|i\rangle$ and $|i+1\rangle$.
Then, the auxiliary forward operator has the form 

\begin{equation}
    \dyad{i+1}{i} = \dyad{x_{n_a}}{x_{n_a}}\otimes \dots\otimes\dyad{\neg x_j }{x_j}\otimes \dots \otimes \dyad{x_1}{x_1}
\end{equation}

where $\neg$ indicates the negation of the bit value. 
This is a tensor product of operators acting on single qubits. 
Then, each term is mapped onto qubit operators using the following equalities:
\begin{equation}
\label{eq:map_proj}
    \begin{aligned}
    \dyad{x}{x}          &= \frac{1}{2}\big(\hat{\mathbf{I}} + (-1)^{x}\hat{\sigma}^z \big) \, , \\
    \dyad{\neg x}{x} &= \frac{1}{2}\big(\hat{\sigma}^x + (-1)^{x}i\hat{\sigma}^y \big) \, , \\
    \end{aligned}
\end{equation}

where $\hat{\sigma}^x$, $\hat{\sigma}^y$ and $\hat{\sigma}^z$ indicate the Pauli operators.
The same considerations apply to the backward auxiliary operator and the projectors.

The last term to consider is the time evolution operator $U(dt)$.
In general, it is not possible to prepare $\hat{U}(dt)$ exactly in an efficient way if the Hamiltonian contains non-commuting terms. 
For this reason we will consider the Trotter-Suzuki approximation of the time evolution operator \cite{Trotter1959,Suzuki1991}.


Having shown how to prepare the Feynamn-Kitaev as a qubit operator, we can now use the Variational Quantum Eigensolver (VQE) \cite{Peruzzo_2014} with such Hamiltonian to optimise the variational parameters of a quantum circuit in order to obtain an approximation of the history state.
Considering a unitary variational circuit ansatz $\ket{\Psi_{\theta}}  = \hat{V}(\theta)\ket{0} $, we will use the quantum computer to measure $E_{\theta} = \bra{\Psi_{\theta}} \hat{C} \ket{\Psi_{\theta}} $ and its derivatives.
Iteratively, a classical processor will determine new values of $\theta$ to minimize $E_{\theta}$.
We remark that minimising $E_{\theta}$ has been proved to be equivalent to the application of McLachlan's variational principle \cite{McLachlan1964,McClean_2013}.
In the following, we will refer to the whole procedure as Variational Feynman Kitaev (VFK) method.

Once a variational approximation of the history state is obtained, we measure expectation values of observables $\hat{O}$ on the system at time step $t$ using the following equality: 

\begin{equation}
\label{eq:obs_eval}
    \expval{\hat{O}(t)} = \bra{\psi_t}\hat{O} \ket{\psi_t} \sim \bra{\Psi_{\theta}} \big[ \hat{O} \otimes \dyad{t}{t} \big] \ket{\Psi_{\theta}} \, .
\end{equation}

An sketch of the method is shown in \cref{fig:vfk_sketch}, while example applications of the final variational history state can be found in \cref{sec:results}.

\subsection{Computational cost on quantum hardware of the Feynman-Kitaev Hamiltonian}
\label{sec:comp_cost}

In this Section we analyze the computational cost of measuring the Feynman-Kitaev Hamiltonian in \cref{eq:clock_hami} on a quantum device.
First, we estimate the number of Pauli strings acting on the clock $n_{\textrm{Pauli}}^{c}$ and on the physical system $n_{\textrm{Pauli}}^{p}$.
Then, the total number of Pauli strings  will be $n_{\textrm{Pauli}}^{\textrm{tot}} \leq n_{\textrm{Pauli}}^{p}* n_{\textrm{Pauli}}^{c}$ .

Given a number of time steps $n$, or $n_a = \log_{2}n$ auxiliary qubits to encode them, we can determine how many Pauli strings will make up the operators acting on the auxiliary system.  
The terms in $C_0$ and $C_1$ are projectors  $\dyad{i}{i}$ with $i \in \{0,\dots,n \}$, can be expressed as $2^{n_a}$ combinations of $\hat{\mathbf{I}}$ and $\hat{\sigma}^z$ as indicated in \cref{eq:map_proj} that can be measured all at once.
On the contrary, $C_2$ contains the forward and backward unitaries, resulting in $n \log_2 n = n_a 2^{n_a}$ different Pauli strings.

For the physical system, we approximated the time evolution operator with a single Trotter-Suzuki step of size $dt$.
The unitary operation corresponding to a Trotter step is not local if the terms of the physical Hamiltonian do not commute, giving a final number of different Pauli strings exponential in the physical qubits, a behaviour that may hinder scalability of the proposed method.
For this reason, we have to resort to a different strategy.
Consider the $m$ sets of non-commuting operators $H_m$ that constitute the physical Hamiltonian $H = \sum_m H_m$.
We can rewrite the operator $C_2$ as

\begin{equation}
    \label{eq:c2_eff}
    \hat{C}_2 = \sum_{i=0}^{n'-1} \sum_{j=0}^{m-1} \hat{U}_j(dt)\otimes \dyad{ij+1}{ij} + \hat{U}_j^{\dagger}(dt) \otimes \dyad{ij}{ij+1}
\end{equation}

where $U_j(dt) = e^{-i H_j dt}$ is the $j$-th set of commuting operators, that can be exactly Trotterized, and $n' = \lfloor\frac{n}{m} \rfloor$.
The physical part of the Feynman-Kitaev Hamiltonian will have $m$ sets of commuting Pauli strings, since all the strings inside the same set can be measured at once. 
This strategy reduces the number of Trotter steps we can encode into $n_a$ auxiliary qubits from $n$ to $n'$, but reduces the computational cost from exponential to polynomial in the number of physical spins. 
In \cref{sec:results} we show how to implement this strategy for the physical system under consideration.

\subsection{Dynamical quantum phase transitions}
\label{sec:DPT}
Given a quantum system governed by the Hamiltonian $\hat{H}$ with initial state  $|\psi(0)\rangle$, the Loschmidt echo is defined by

\begin{equation}
\label{eq:loschmidt_echo}
    L(t) \equiv \abs{\bra{\psi(0)} e^{-i \hat{H} t} \ket{\psi(0)}}^{2} = \abs{\braket{\psi(0)}{\psi(t)}}^{2} \, .
\end{equation}

Loschmidt echoes are important quantities in many-body theory and appear in various contexts such as quantum chaos \cite{peres1984_pra,gorin2006_jpr} or the Schwinger's particle production mechanism \cite{schwinger1950_pr,martinez2016_nat}.
In particular, we will consider the measurement of Loschmidt echoes for the identification of dynamical quantum phase transitions (DQPT) \cite{heyl2013_dpt_tfim,heyl2018_dpt_review}.
Because the variational history state encodes the superposition of all different time-evolved physical states $\ket{\psi(t)}$ we will show in the following that it is computationally cheap to evaluate the Locshmidt echo on it.

Dynamical phase transitions can be identified by a nonanalytic behavior of the Lochsmidt echo and of its rate function $\lambda(t)$

\begin{equation}
   \label{eq:rate_fun}
    \lambda (t) = - \lim_{D \rightarrow \infty} \frac{1}{D} \log \left[ L(t) \right]
\end{equation}
where $D$ is the number of degrees of freedom in our quantum system.
Such phase transitions have been identified in many quantum systems \cite{heyl2018_dpt_review} and also observed in experiments performed on analog quantum simulators \cite{jurevic2017_prl,flaeschner2018_np,xinfang2020_prl}.

In the following, we will show how to use the VFK method to study these phenomena.
Assuming that the variational parameters of a circuit approximating the history state are obtained, we use an additional qubit to perform an Hadamard test \cite{ekert2002_prl} and measure the real and the imaginary part of $\braket{\psi(t_i)}{\psi(t_j)}$ in order to calculate the Loschmidt echo.
This only requires an additional qubit and the gate overhead required to calculate the Loschmidt echo with VFK state is negligible with respect to the total number of gates of the variational state.

A more detailed explanation of the Hadamard test, as well as the complete circuit, can be found in \cref{sec:losch_echo}, while an application of this method is reported in \cref{sec:results}.

\section{Results}
\label{sec:results}

To demonstrate the viability of the VFK method, we consider the Transverse Field Ising Model on an open 1-dimensional chain, 

\begin{equation}
\label{eq:tfim_hami}
\hat{H}= J \sum_{i=0}^{n_{s}-1} \hat{\sigma}^{z}_{i}\hat{\sigma}^{z}_{i+1} + h\sum_{i=0}^{n_{s}} \hat{\sigma}^{x}_{i} \quad .
\end{equation}

The first term accounts for interaction between spins while the latter represents a local and uniform magnetic field along the transverse direction $x$. 
For our simulations, we considered $J=\frac{1}{4}$, $h=1$.
We use a classical computer to simulate the quantum hardware in an ideal case, without noise sources and with full access to the state vector produced by the quantum circuit (state-vector simulations).
The simulations have been performed both in Python using IBM’s open-source library for quantum computing, \textrm{Qiskit} \cite{Qiskit}, and in Julia \cite{bezanson2017julia} with the \textrm{Yao.jl} \cite{Luo_2020} and PastaQ \cite{pastaq} frameworks.
The code can be found at \cite{barison2022github,barison2022mc}.

We considered a variational ansatz of the form 

\begin{equation}
    \label{eq:ansatz}
    \hat{V}(\theta) = \prod_{l=1}^{d} \Bigg[  \prod_{\substack{a \in A}} \bigg( \prod_{\substack{i \neq j \\ i,j \in P}} \hat{G}_{ij,l} \prod_{\substack{p \in P}} \hat{G}_{ap,l} \bigg) \Bigg]
\end{equation}

where we denoted the set of qubits encoding the physical system $P$ and the set of auxiliary qubits $A$, while $G_{ij,l}$ is a parameterised gate of the form

\begin{figure}[h!]
    \centerline{
\Qcircuit @C=1em @R=1.5em {
& \multigate{1}{\hat{G}} & \qw \\
& \ghost{G}        & \qw
} ~~~~=~~~~ 
\Qcircuit @C=1em @R=1em {
& \gate{\hat{R}_{x}(\theta_1)}  & \gate{\hat{R}_{y}(\theta_2)} & \ctrl{1} & \qw \\ 
& \gate{\hat{R}_{x}(\theta_3)}  & \gate{ \hat{R}_{y}(\theta_4)} & \targ   & \qw   
}
}
    \label{fig:g_gate}
\end{figure}

The total number of blocks, or depth, is $d$. 
The total number of variational parameters is then given by

\begin{equation}
    \label{eq:num-parameters-ansatz}
    n_{pars} = 2 d  n_{a} (n_{s}^{2} + n_{s})
\end{equation}

where $n_s$ and $n_a$ indicate the number of spin and auxiliary qubits, respectively.

The Ising Hamiltonian in \cref{eq:tfim_hami} has two sets of non commuting operators $H_X = \sum_{i=0}^{n_{s}} \hat{\sigma}^{x}_{i} $ and $H_{ZZ} = \sum_{i=0}^{n_{s}-1} \hat{\sigma}^{z}_{i}\hat{\sigma}^{z}_{i+1}$. 
We make use of the strategy presented in \cref{sec:comp_cost} in order to have a polynomial cost.
We build the Feynman-Kitaev Hamiltonian by applying $U_{X}(dt) = \exp[-i H_X dt]$ at even clock values and $U_{ZZ}(dt) = \exp[-i H_{ZZ} dt]$ at the odd ones.
The new amount of Trotter step that can be encoded in $n_a$ auxiliary qubits will be $n' = \frac{n}{2}$.

\begin{figure}[h]
        \vspace{2mm}
        \includegraphics[width=1.0\columnwidth]{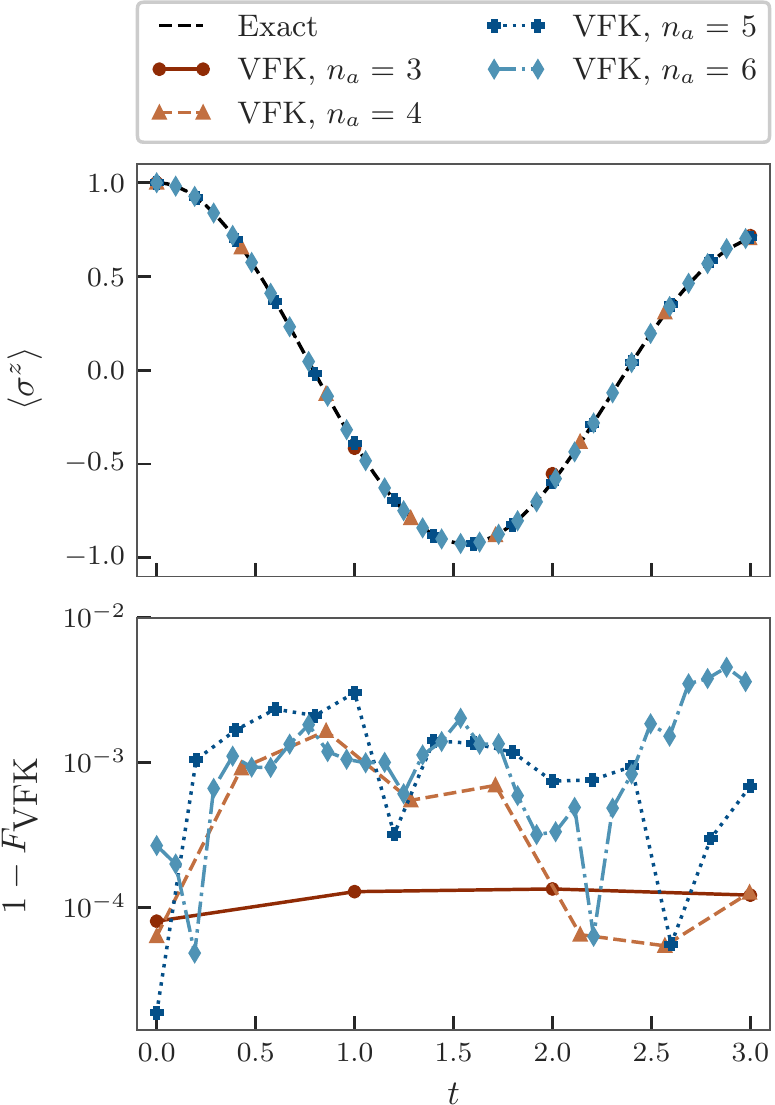} 
        \caption{Magnetization (top) and infidelities (bottom) measured on the VFK history state. We considered a system of $n_s = 6$ spins initialised in $|0\rangle ^{\otimes 6}$ and different auxiliary qubits $n_a$ for a simulation time $T_{e}= 3.0$. $F_{\text{VFK}}$ at time $t_i$ indicates the fidelity measured between the Trotter-Suzuki state with $i$ steps and the VFK state projected on the $| i \rangle$ subspace.}
        \label{fig:obs-fid_comp}
\end{figure}

First, we considered a system with $n_s = 6$ spins and up to $n_a = 6$ auxiliary qubits for a total simulation time $T_e = 3$.
We used the variational wavefunctions obtained with VFK to estimate expectation values of physical quantities as indicated in \cref{eq:obs_eval}.
In particular, we evaluated the total magnetization along the $z$ axis
\begin{equation}
    \sigma^z = \frac{1}{n_s} \sum_{i=1}^{n_s} \sigma^{z}_i
\end{equation}
through the entire time evolution.
To assess the accuracy of these variational history states obtained with the VFK method, we measured the infidelity $1- F_{\text{VFK}}$ with respect to Trotter-Suzuki wavefunctions.
For every simulated time $t_i$, we prepared the Trotter-Suzuki state $|\psi_{i}^{TS}\rangle = \hat{U}^{i}_{TS}(dt)|\psi(0)\rangle$ with $dt = \frac{T_e}{2^{n_a-1}}$ and evaluated the infidelity with VFK history state projected onto the $|i\rangle$ subspace.
Results are reported in \cref{fig:obs-fid_comp}.

Increasing the number of auxiliary qubits while keeping the total simulation time $T_e$ fixed decreases the size of a single Trotter step, decreasing also the error with respect to the exact state.
We remark that, if the variational ansatz is able to reproduce exactly the history state, the variational history state can be as accurate as the corresponding Trotter-Suzuki approximation.


\begin{figure*}[ht]
       \includegraphics[width=1.0\textwidth]{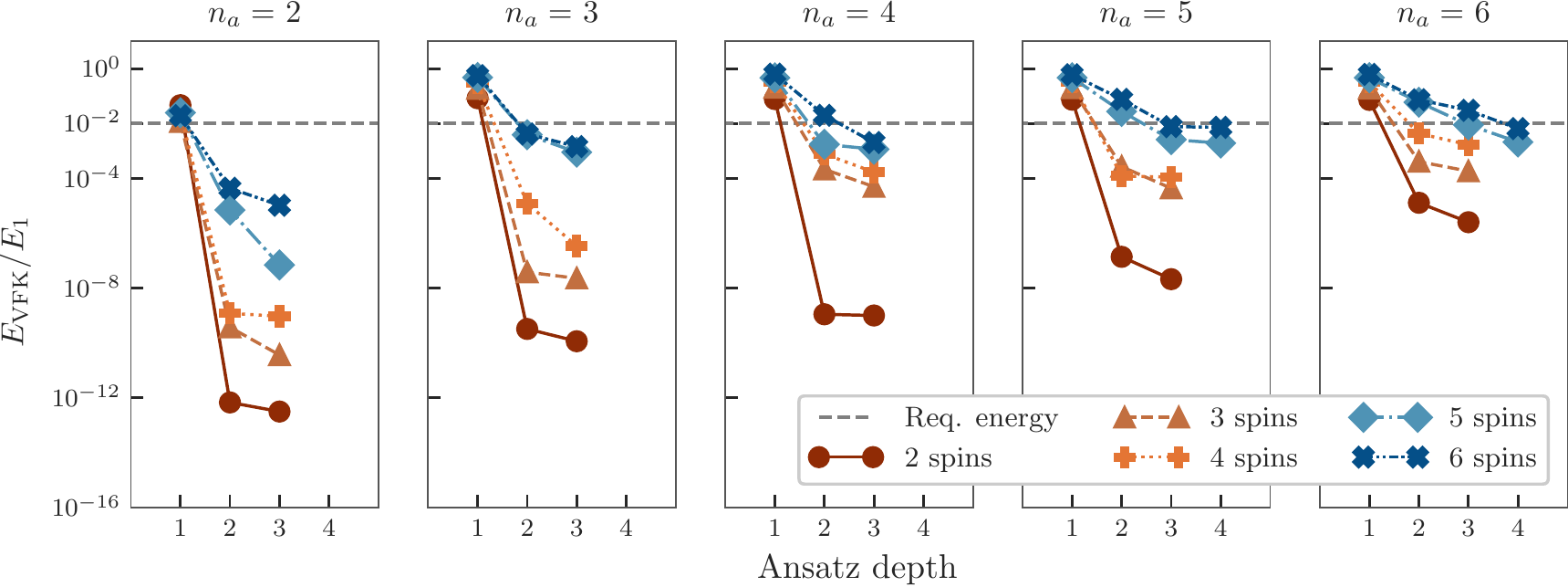} 
        \caption{Final energy obtained with the VFK algorithm ($E_{\text{VFK}}$) over the energy of the first excited state ($E_1$) obtained using a classical simulator. The plot shows the final energy obtained for a system of $n_s$ spins and $2^{n_a-1}$ time steps as a function of the depth of ansatz in \cref{eq:ansatz}. The gray, dashed lines indicate the energies we require to consider the VFK converged. We fixed the simulation time $T_{e}= 3.0$, therefore increasing the number of auxiliary qubits reduces the time step and the Trotter-Suzuki approximation error, accordingly. The corresponding fidelities can be found in \cref{sec:fidelities_vqe}}
        \label{fig:depth_comp}
\end{figure*}

We run multiple simulations for varying system and clock sizes $n_s,n_a \in \left[2,3,4,5,6 \right]$.
The ground-state optimisation is performed using the Variational Quantum Eigensolver (VQE) algorithm and the \textrm{ADAM} optimiser \cite{kingma2014adam}. 
The time evolution unitaries in the target Feynman-Kitaev Hamiltonian are smoothly varied from $\hat{U}(dt=0)$ to $\hat{U}(dt=T_e/2^{n_a-1})$ in order to help the convergence of the method (see \cref{sec:method_opt} for additional details).

During the optimisation we monitor the energy of the state, which we showed to bound the infidelity with respect to the target state.
Once it gets below a certain threshold, we stop the optimisation and consider that we have reached convergence.
Since the energy of the first excited states of the Feynman-Kitaev Hamiltonian decreases as the number of auxiliary qubits increases, the threshold is chosen to be $10^{-2} \cdot E_1$, where $E_1$ indicates the energy of the first excited state. 
As indicated in \cite{kitaev02_qc},  $E_1$ can be derived analytically from an unitarily equivalent Hamiltonian and its value is $E_1 = 1- \cos(\pi /2^{n_a})$.
For this reason, we are able to use $E_1$ as convergence threshold.

The complexity of the optimisation procedure increases with the number of auxiliary qubits, however we are able to reach convergence by increasing the ansatz depth $d$.
This is consistent with the fact that preparing  the \textit{exact history state} would require an exponentially deep circuit with the clock system size.
In \cref{fig:depth_comp} we report the final energies obtained as a function of the ansatz depth.
For comparison, in \cref{sec:fidelities_vqe} we report the infidelities with respect to the exact ground state for the same calculations.

\begin{figure}[h]
        \includegraphics[width=1.0\columnwidth]{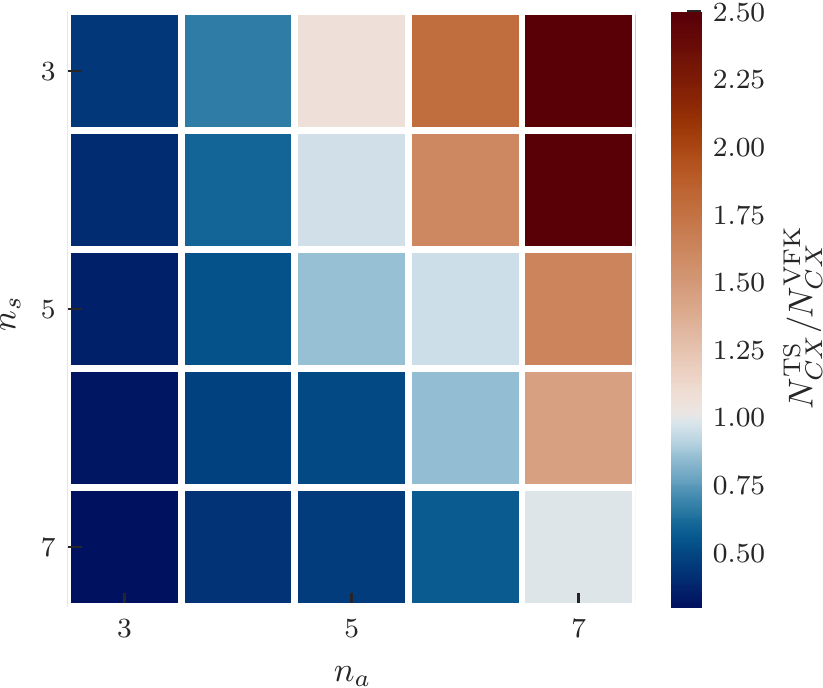} 
        \caption{Comparison between two-qubits gates required by the Trotter-Suzuki circuit ($N^{TS}_{\text{CX}}$) and by the variational ansatz ($N^{\text{VFK}}_{\text{CX}}$), given the same system size and Trotter steps. $N^{\text{VFK}}_{\text{CX}}$ corresponds to the minimal circuit required to make VFK converge and depends on ansatz type and depth.
        $N^{TS}_{\text{CX}}$ is fixed, given system size and number of steps. 
        We considered the ansatz in \cref{eq:ansatz} and a simulation time $T_{e}= 3.0$. }
        \label{fig:gate_comp}
\end{figure}

Then, we considered the smallest depth necessary to reach convergence, and compared the number of two-qubit gates needed by the variational-circuit with the number of two-qubit gates needed by a standard Trotter-Suzuki circuit with the same number of time-steps. 
The results are indicated in \cref{fig:gate_comp}.
We only report the comparison for two qubit gates and not for one-qubit gates because the latters are usually not a bottleneck in NISQ devices.

In \cref{fig:gate_comp} we show that the VFK history state has an overhead compared to Trotter if few time-steps are considered, but as the number of time-steps is increased, VFK method has a favorable scaling.
Moreover, the VFK history state contains more information than the state obtained at the end of the Trotter evolution, which can be used, for example, to compute Lochsmidt echoes.

In this regard, we show that the VFK variational wavefunction can be used to study dynamical quantum phase transitions.
We still consider the Transverse-Field Ising Model with $J=1/4$ and $h=1$. 
As the equilibrium model exhibits a quantum phase-transition in the ground-state when going from $h<J$ to $h>J$, a dynamical quantum phase transition is known to appear if the system is quenched from one region to the other \cite{heyl2018_dpt_review}.

\begin{figure}[h]
        \includegraphics[width=1.0\columnwidth]{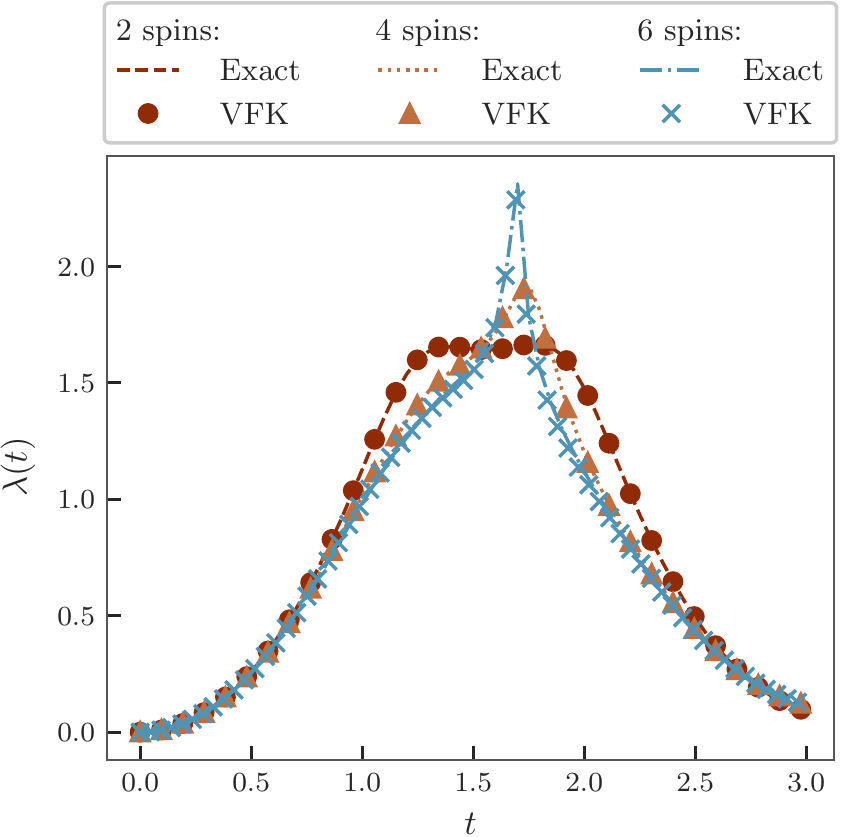} 
        \caption{Rate function of the Loschmidt echo in the Transverse Field Ising Model measured using the VFK variational state. The plot shows  a dynamical phase transition for an open chain initialised in the $|0\rangle^{\otimes n_s}$ state and evolved under the Hamiltonian in \cref{eq:tfim_hami}.
        The nonanalyticity appears when the number of spin is increased.
        We considered a number of auxiliary qubits $n_a = 6$ for $n_s = 2,4$ and $n_a =7 $ for $n_s =6$.}
        \label{fig:DPT_comp}
\end{figure}

We consider a system initialised at $|\psi(0)\rangle = |0\rangle^{\otimes n_s}$, the ground-state of the Hamiltonian for $h=0$, and we then compute the VFK history state using the same procedure described before for $T_e=3$, $h=1$ and varying system sizes.
After determining the optimal parameters, we used the Hadamard test to evaluate the rate function of the Loschmidt echo as described in \cref{sec:losch_echo}.
In \cref{fig:DPT_comp} we show that a nonanalytic behaviour appears as the system size grows.
We are able to improve the time-resolution by increasing the number of auxiliary qubits.
As the number of spins approaches the thermodynamic limit, the qubit overhead required to characterise the cusp becomes negligible.

Finally, we investigate the algorithm performance in the presence of noise.
We follow the noise model proposed by Kandala, Mezzacapo et al. in \cite{Kandala2017}, where the effects of decoherence are approximated by  an amplitude damping and dephasing channel acting  after each \textit{fundamental} gate on the system density matrix $\rho$ as

\begin{equation}
    \begin{aligned}
        \rho & \rightarrow E^{a}_0 \rho E^{a \dagger}_0 + E^{a}_1 \rho E^{a \dagger}_1 \, ,\\
        \rho & \rightarrow E^{d}_0 \rho E^{d \dagger}_0 + E^{d}_1 \rho E^{d \dagger}_1
    \end{aligned}
\end{equation}

with 

\begin{equation}
    E^{a}_0 = \begin{bmatrix} 
	1 & 0 \\
	0 & \sqrt{1-p^a} \\
	\end{bmatrix} \, , \,
	E^{a}_1 = \begin{bmatrix} 
	0 & \sqrt{p^a} \\
	0 & 0 \\
	\end{bmatrix} 
\end{equation}

and
\begin{equation}
    E^{d}_0 = \begin{bmatrix} 
	1 & 0 \\
	0 & \sqrt{1-p^d} \\
	\end{bmatrix} \, , \,
	E^{d}_1 = \begin{bmatrix} 
	0 & 0 \\
	0 & \sqrt{p^d} \\
	\end{bmatrix} \, .
\end{equation}
The fundamental gates are taken to be CNOT, $\hat{R}_x$, $\hat{R}_y$ and $\hat{R}_z$; all other gates, such as the $\hat{R}_{zz}$ used in the Trotter circuit, are decomposed into a sequence of those gates.
The 1- and 2- qubit error rates $p^{a,d}_1$ and $p^{a,d}_2$ are estimated from QPU data such as the relaxation and dephasing time available at \cite{ibm_services,ionq_services}.
More details on the estimation of the error rates can be found in \cref{sec:noise_model}.

\begin{figure}[h]
        \includegraphics[width=1.0\columnwidth]{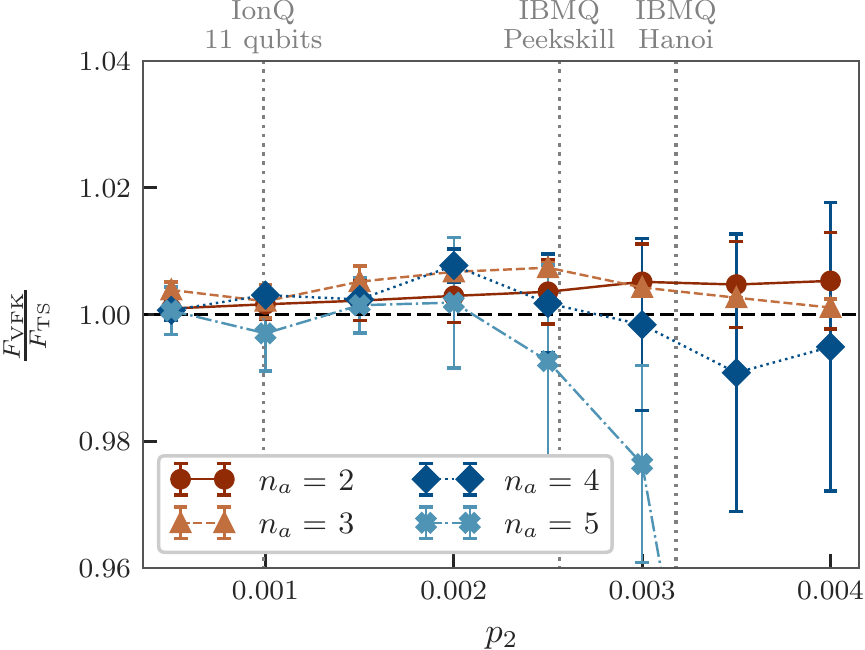} 
        \caption{Comparison between the mean infidelity of noisy VFK and Trotter-Suzuki circuits through a time evolution to $T_e = 3.0$.
        Markers indicate the mean value, while the bars the standard deviation over a time evolution.
        We fixed the error probability for single qubit gates $p_1 = 2 \cdot 10^{-4}$.
        We considered a system with $n_s=2$ and different number of auxiliary qubits $n_a$, Trotter steps are varied accordingly.
        The horizontal line indicates equal mean fidelity between VFK and Trotter-Suzuki circuits.
        Vertical lines indicate the mean $p_2$ error probability for reference devices.
        }
        \label{fig:noise_comp}
\end{figure}

In \cref{fig:noise_comp} we compare the mean fidelity $F_\text{VFK}$ obtained from the VFK calculation on noisy system of $n_s = 2$ spins and different $n_a$ against the mean fidelity of the trotter evolution $F_\text{TS}$ (we take the fidelity to be $F(\rho, \sigma) = \left[\textrm{Tr}\left(\sqrt{\sqrt{\rho}\sigma \sqrt{\rho}} \right) \right]^2$, where $\rho, \sigma$ are two density matrices).
We study how their ratio $F_\text{VFK}/F_\text{TS}$ varies as a function of different 2-qubit error rates $p^a_2 = p^d_2 = p_2$, while keeping the single qubit error rate constant to an average value $p^a_1 = p^d_1 = p_1 = 2 \cdot 10^{-4}$ that is consistent with real-world devices. 
This approach is motivated by the fact that today's devices are mainly limited in the fidelity they can achieve when performing 2-qubit operations.

The plot indicates comparable performances between the two methods, with an advantage of VFK for $n_a =2,3$.
As the number of auxiliary qubits is increased to $n_a =4,5$, the VFK optimisation must converge towards smaller values of the energy in order to give sensible results. However, at higher error rates we were not able to succesfully optimise the variational parameters, which can be seen in the figure by the fact that the performance of the VFK method then degrades.
To give a better idea of where today's QPUs stand, some vertical lines represent the mean values of $p_2$ for commercially-available devices, which shows that Ion-based setups have a low-enough error rate and could be able to use the VFK algorithm efficiently already today.


\section{Conclusions and outlook}
\label{sec:discussion}

In this work we showed a strategy to study dynamical properties of a quantum system using variational ground-state quantum algorithms. 
Our method, which we called variational-Feynman Kitaev (VFK), combines the Feynman-Kitaev Hamiltonian together with existing variational-ground state methods in order to study dynamical systems with the limited resources available on today's NISQ devices.

First, we showed how the Feynman-Kitaev Hamiltonian can be efficiently implemented on a quantum computer using a binary encoding for the auxiliary system and the Trotter-Suzuki approximation for the time evolution operator.
Then, we demonstrated our method studying the transverse field Ising model and showing that it scales favorably compared to a Trotter-Suzuki evolution when many time-steps are needed.
We investigated the convergence of the algorithm as the number of spins and auxiliary qubits increases and compared the final circuits with the corresponding Trotter-Suzuki approximation.
Considering a noise model accounting for decoherence, we assessed that our algorithm yields slightly better results than a plain Trotter evolution when the two qubits error rate is not too detrimental.
As analysed in \cite{McClean_2016}, variational methods are robust against certain types of quantum errors, a promising feature for future hardware demonstrations.
Finally, we showed that it is possible to exploit the structure of the VFK history state to measure dynamical quantities such as the Loschmidt echo efficiently, and gave an example of how to detect dynamical quantum phase transitions in the Transverse Field Ising model.

As a future research direction, we believe that it would be of interest to investigate the application of the VFK method to other physical systems, as well as its application to estimate quantities that are not easily accessible via standard Trotterisation. 
Similarly to all other variational algorithms, the choice of the right ansatz is fundamental for the algorithm to succeed.
In the ansatz we proposed, the gates required between auxiliary and system qubits currently limit the implementation on hardware.
However, we envision application of the method in devices with high qubit connectivity.
Moreover, we showed that once the variational history state is obtained, it has applications beyond the evaluation of  expectation values of time-dependent observables.

\section*{Data availability}
The code used to run the simulations is open source and can be found at \cite{barison2022github,barison2022mc}.

\section*{Acknowledgments}
This research was supported by the NCCR MARVEL, a National Centre of Competence in Research, funded by the Swiss National Science Foundation (grant number 205602).

\onecolumngrid
\appendix


\section{Comparison with projected - Variational Quantum Dynamics}
\label{sec:com_pvqd}


In this Appendix we compare the VFK method to the p-VQD method presented in \cite{barison2021efficient}.
We consider the same TFIM in \cref{eq:tfim_hami} and the same parameters.
The p-VQD variational state is prepared using the ansatz presented in \cite{barison2021efficient}. 
We define the mean integrated infidelity

\begin{equation}
    \delta_{F}(T_e) = \frac{1}{T_e} \int_{t = 0}^{T_e} \left( 1 - |\braket{\psi_{TS}(t)}{\psi_{\theta_t}}|^{2} \right) dt = \frac{\Delta_{F}(T_e)}{T_e}
\end{equation}

where $|\psi_{TS}(t) \rangle$ is the Trotter-Suzuki state prepared at time  $t$, while $|\psi_{\theta_t} \rangle$ indicates the variational state that approximates it.
In the VFK method, the variational history state $|\Psi_{\theta}\rangle $ is prepared and projected onto the $|t\rangle$ subspace before calculating the fidelity.
We fix a total simulation time $T_e$ and the number of auxiliary qubits in the VFK method determine the number of time steps for both methods.
Given that, differently from VFK,  the variational wavefunction prepared by the p-VQD algorithm encodes the evolution of the quantum state only at certain time $t$, the ans\"atze used in the two methods are different.
For VFK, the ansatz is the same presented  in \cref{eq:ansatz}, with depth of converged calculations in \cref{fig:depth_comp}.
For p-VQD, we use the same ansatz presented in \cite{barison2021efficient} with depth $d=3$, resulting in a total number of $d*(n_s+(n_s-1)) +n_s = 7 n_s - 3$ free parameters.

\begin{figure*}[h]
        \includegraphics[width=1.0\textwidth]{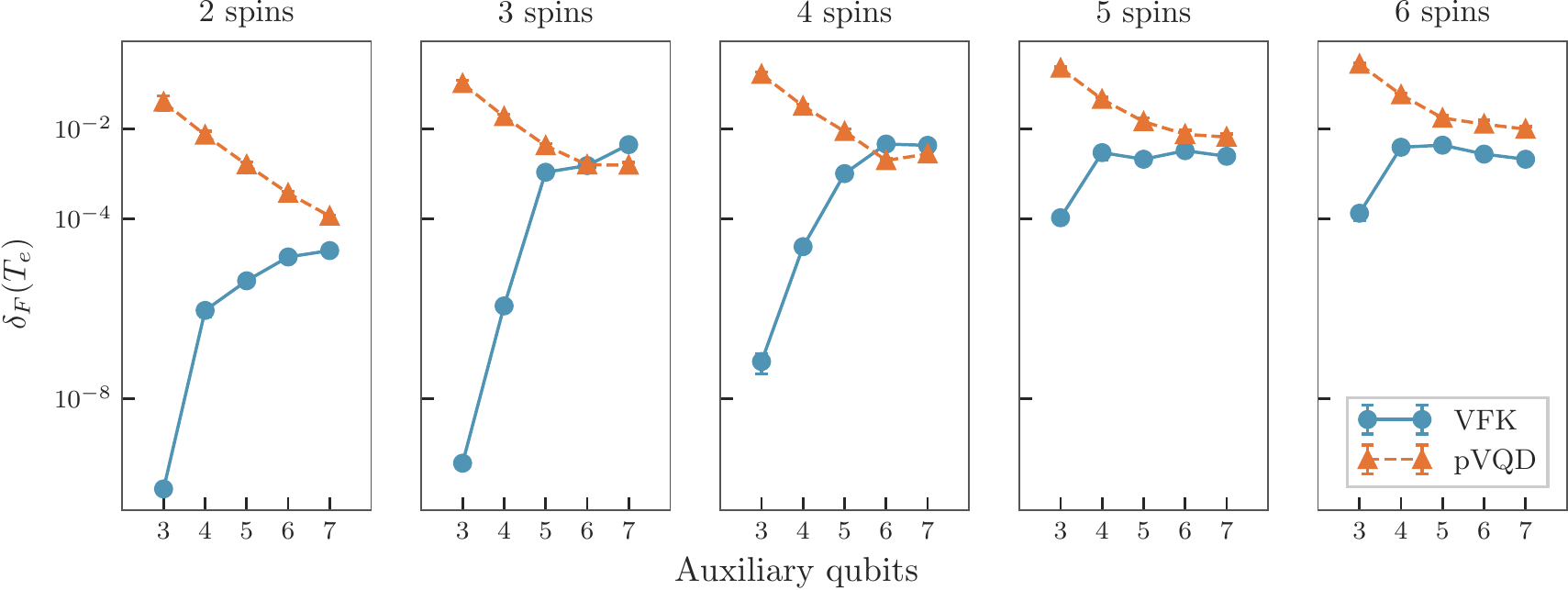} 
        \caption{Mean infidelity over an entire time evolution as a function of the number of spins and auxiliary qubits. The plot compares the mean infidelity between the variational state obtained with VFK (after projecting to $|t\rangle$ subspace) or p-VQD and the exact state.
        For the VQE we used the ansatz in \cref{eq:ansatz}, while for p-VQD the alternating rotations ansatz presented in \cite{barison2021efficient}. We considered a total simulation time $T_{e}= 3.0$ and a number of optimisation steps comparable between the two methods. The analysis is performed using a state-vector simulator and the ADAM optimiser for both methods. }
        \label{fig:fid_comp}
\end{figure*}

We use the ADAM optimiser and a comparable number of optimisation steps.
Results of this comparison are shown in \cref{fig:fid_comp}.

When the number of time steps is small, the infidelity of the VFK method is lower.
If we choose a time step small enough we are able to improve the accuracy of the p-VQD method, thanks to parameter optimisation at every time step.
Provided a converged optimization, they are expected to converge to the same mean integrated infidelity in the limit of infinite time steps.

\section{Optimisation strategy}
\label{sec:method_opt}

As a variational algorithm, the optimisation of the wavefunction  using VQE can be hindered by local minima or the presence of barren plateaus \cite{McClean_2018,Grant_2019,Cerezo_2021}. 
In particular, the definition of the cost function in terms of global observables leads to the emergence of such plateaus even when $V(\theta)$ is a  shallow circuit \cite{Cerezo_2021}.
To improve the optimisation procedure, we devise a specific strategy.
Consider $dt =0$, in this case the history state has the form

\begin{equation}
\label{eq:init_state}
    \ket{\Psi_0} = \frac{1}{\sqrt{n+1}}\sum_{i=0}^{n}\ket{\psi(0)} \ket{i} \, .
\end{equation}

When $dt$ is infinitesimal, the history state $\ket{\Psi}$ is infinitesimally close to $\ket{\Psi_0}$. 
Assuming $\ket{\psi(0)}$ is a simple state to prepare or, equivalently, that we have an efficient approximation to it, we start with the circuit in the state $\ket{\Psi_0}$ and initialize the variational gates as identity operators. 
Then, we substitute every $\hat{U}(dt)$ in the Feynman-Kitaev Hamiltonian with its $k$-root $\sqrt[k]{\hat{U}(dt)} = \hat{U}(dt/k)$.
Considering $k$ large enough, we can start the optimisation close to the optimum.
Finally, we  optimise the wavefunction for decreasing $k$ until $k=1$ and the targeted clock Hamiltonian is obtained.
In general, every function $f(dt,k)$ can be considered and the Feynman-Kitaev Hamiltonian built upon $\hat{U}[f(dt,k)]$, e.g. $f(dt,k) = dt \frac{k}{k_0}$ with $k \in \{1,\dots,k_0 \}$ and $k_0$ an arbitrary large number.

\section{Fidelity of the VQE calculations}
\label{sec:fidelities_vqe}

In this Appendix we show the fidelity between the variational state obtained with the VFK algorithm  and  the corresponding exact history state.
As explained in \cref{sec:feynman-kitaev-hamiltonian}, the fidelity is upper bounded by the normalized energy.
Indeed, given a state $\ket{\phi} = \sqrt{1-\epsilon^2}\ket{\Psi} + \epsilon\ket{\delta}$ close to the target state $\ket{\Psi}$, where $\epsilon\approx 0,\, \, \braket{\Psi}{\delta} = 0 $, we can calculate its fidelity and energy as a function of $\epsilon$.

\begin{figure*}[h]
        \includegraphics[width=1.0\textwidth]{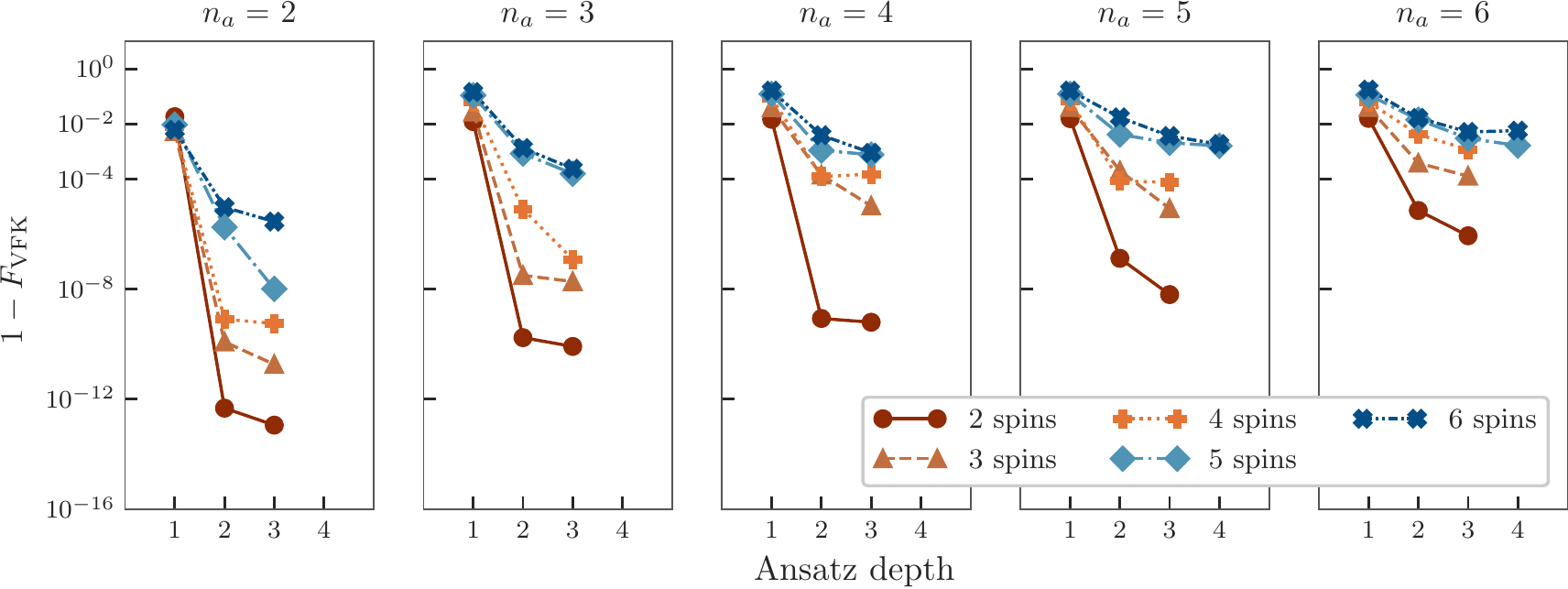} 
        \caption{Final infidelities of the states obtained with the VFK algorithm ($1-F_{\text{VFK}}$) using a classical simulator with respect to the exact history state. The plot report the infidelities corresponding to the calculations presented in \cref{fig:depth_comp}.}
        \label{fig:fid_vqe_comp}
\end{figure*}

In particular, we obtain that $1-F(\phi, \Psi) = \epsilon^2$ and $E(\ket{\phi}) = \epsilon^2 E(\ket{\delta})$, where we defined $F(\phi, \Psi) =  |\langle \phi | \Psi \rangle |^2$ and to calculate $E(\ket{\phi})$ we considered that $E(\ket{\Psi}) = 0$ by construction.
Given that $\ket{\delta}$ is orthogonal to $\ket{\Psi}$, $E(\ket{\delta}) \geq E_1$, where $E_1$ is the energy of the first excited state.
For this reason, we can conclude that the infidelity is upper bounded by the energy normalized by $E_1$, which can be calculated analytically, as shown in \cite{kitaev02_qc}. 
In \cref{fig:fid_vqe_comp} we report the infidelities for states whose energy is shown in \cref{fig:depth_comp}

\section{Measuring Loschmidt echoes}
\label{sec:losch_echo}

In this Appendix, we illustrate how to extend the VFK circuit in order to measure Loschmidt echoes.
Consider the overlap between the system quantum state evolved at two different times $t_i$ and $t_j$, namely $\braket{\psi(t_i)}{\psi(t_j)}$.
In general, this quantity is complex, therefore we have to perform an Hadamard test to evaluate its real and imaginary part.
In the following, we describe how to measure the real part; the same procedure applies for the imaginary part, with a slight modification. 

First, we define the time-swap operator as

\begin{equation}
    \hat{\text{T}}_{\text{SWAP}}(t_i,t_j) = \bigotimes_{k \in T_{ij}} \hat{\sigma}^x_k \, ,
\end{equation}

where $T_{ij}$ indicates the set of bits that differs in the bit string encoding of $t_i$ and $t_j$.
We use the set of parameters obtained with the VFK method to initialise the variational circuit $V(\theta)$.
We control the action of the time-swap operator over the clock system using an additional auxiliary qubit initialized in $|+\rangle = \frac{|0\rangle + |1\rangle}{\sqrt{2}} = H|0\rangle$.
In the following, we will refer to the qubits of the auxiliary clock system as the clock qubits, to distinguish them from the auxiliary qubit added to perform the Hadamard test.
Then, we apply again an Hadamard gate to the auxiliary qubit and we measure the auxiliary and the clock qubits.
We post-select the measurements in which the measurement of the clock system returns the binary string corresponding to $t_i$ (or $t_j$).
We indicate the number of post-selected measurements with the auxiliary qubit in $0$ ($1$) as $N_0$ ($N_1$).

Finally , the real part of the overlap will be

\begin{equation}
    \braket{\psi(t_i)}{\psi(t_j)} = 2^{n_a} \frac{N_0 - N_1}{N_{\text{shots}}}
\end{equation}

where $N_{\text{shots}}$ indicates the number of times the system has been prepared and measured, while the prefactor $2^{n_a}$ is required for the normalisation of the history state.
Considered a system with $2$ system qubits and a single clock qubit, the final circuit to measure the real part of $\braket{\psi(t_0)}{\psi(t_1)}$ has the form

\begin{figure}[h]
    \centerline{
\Qcircuit @C=1em @R=1.5em {
    \lstick{\ket{0}}  & \gate{H}                   & \ctrl{1} & \gate{H} & \meter  \\
    \lstick{A}        & \multigate{2}{V(\theta)}   & \targ    & \qw      & \meter  \\
                      & \ghost{V(\theta)}          & \qw      & \qw      & \qw \\
                      & \ghost{V(\theta)}          & \qw      & \qw      & \qw
   \inputgroupv{3}{4}{.8em}{.8em}{P}
        }
    }   
\label{fig:losch_circ}
\end{figure}

where $P$ indicates the two system qubits and $A$ the clock qubit.
To measure the imaginary part of the overlap, the same procedure applies, with the auxiliary qubit initialised in $ \frac{|0\rangle -i |1\rangle}{\sqrt{2}} = HS^{\dagger}|0\rangle$.

\section{Noise model for the numerical experiments}
\label{sec:noise_model}

In this Appendix, we describe how we included decoherence effects in the numerical simulations shown in \cref{fig:noise_comp}.
As indicated in the main text, we followed the noise model proposed by Kandala et al. in \cite{Kandala2017}, where the effects of decoherence are approximated by the successive application of an amplitude damping and dephasing channel acting  after each gate on the system density matrix 

\begin{equation}
    \begin{aligned}
        \rho & \rightarrow E^{a}_0 \rho E^{a \dagger}_0 + E^{a}_1 \rho E^{a \dagger}_1 \, ,\\
        \rho & \rightarrow E^{d}_0 \rho E^{d \dagger}_0 + E^{d}_1 \rho E^{d \dagger}_1
    \end{aligned}
\end{equation}

with 

\begin{equation}
    E^{a}_0 = \begin{bmatrix} 
	1 & 0 \\
	0 & \sqrt{1-p^a} \\
	\end{bmatrix} \, , \,
	E^{a}_1 = \begin{bmatrix} 
	0 & \sqrt{p^a} \\
	0 & 0 \\
	\end{bmatrix} 
	\, \, \, \, \, \, \text{and} \, \, \, \, \, \,
	E^{d}_0 = \begin{bmatrix} 
	1 & 0 \\
	0 & \sqrt{1-p^d} \\
	\end{bmatrix} \, , \,
	E^{d}_1 = \begin{bmatrix} 
	0 & 0 \\
	0 & \sqrt{p^d} \\
	\end{bmatrix} \, .
\end{equation}

The error rates $p^a$ and $p^d$ can be determined from actual device data.
In particular, we have

\begin{equation}
    p^a = 1- e^{-\tau/T_1} \, \, \, , \, \, \, p^d_{g} = 1- e^{-2\tau/T_{\phi}}
\end{equation}

where $T_1$ is the relaxation time of the qubit, $\tau$ is the gate time and $T_\phi = 2T_1 T_2/(2T_1 - T_2)$, given $T_2$ the coherence time of the qubit.
When the time of a single- or two- qubit gate is substituted in $\tau$, we obtain $p^{a,d}_1$ and $p^{a,d}_2$, respectively.
We used data found in  \cite{ibm_services,ionq_services} to estimate an average value of $p^{a,d}_1$ and a range of $p^{a,d}_2$ to analyze in our simulations.
Finally, in order to simplify the comparison in \cref{fig:noise_comp}, we considered $p^a = p^d$ for both single- or two- qubit gates.
This approximation holds when $T_1 \sim T_2$, which is true on an average basis for superconducting devices \cite{ibm_services}.

For completeness, we report in \cref{tab:noise_data} the data of the devices indicated in \cref{fig:noise_comp}.

\begin{table}[h]
    \centering
    \begin{tabular}{|c|c|c|c|c|}
        \hline
        Device & $T_1$ $(\mu s)$ & $T_2$ $(\mu s)$& $\tau_{2}$ $(\mu s)$& $\tau_{1}$ $(\mu s)$\\
        \hline
        \hline
        IBMQ Peekskill & 250.12 & 242.82 & 0.420571 & 0.0035  \\
        \hline
        IBMQ Hanoi & 160.64 & 144.7 & 0.318603 & 0.0035  \\
        \hline
        IonQ 11 qubits & $>10^7$ & $2*10^5$ & 210 & 10  \\
        \hline
    \end{tabular}
    \caption{Average coherence values and gate times of the devices indicated in \cref{fig:noise_comp}, as of March 2022. }
    \label{tab:noise_data}
\end{table}

\twocolumngrid

\end{document}